\documentclass[twocolumn,showpacs,fleqn,nobibnotes]{revtex4}

\usepackage{amsmath}
\usepackage{graphicx}
\usepackage{float}
\usepackage{subfigure}

\def\lsim{\raise0.3ex\hbox{$<$\kern-0.75em\raise-1.1ex\hbox{$\sim$}}}
\def\gsim{\raise0.3ex\hbox{$>$\kern-0.75em\raise-1.1ex\hbox{$\sim$}}}

\begin{document}
\newcommand\ie {{\it i.e.}}
\newcommand\eg {{\it e.g.}}
\newcommand\etc{{\it etc.}}
\newcommand\cf {{\it cf.}}
\newcommand\etal {{\it et al.}}
\newcommand{\be}{\begin{eqnarray}}
\newcommand{\ee}{\end{eqnarray}}
\newcommand{\jp}{$ J/ \psi $}
\newcommand{\pp}{$ \psi^{ \prime} $}
\newcommand{\ppp}{$ \psi^{ \prime \prime } $}
\newcommand{\dd}[2]{$ #1 \overline #2 $}
\newcommand\noi {\noindent}

\title{Enhancement of prompt photons in ultrarelativistic proton-proton collisions from nonlinear gluon evolution at small-$x$}
\author{C. Brenner Mariotto $^{a}$ and V.P. Gon\c{c}alves
$^{b}$}

\affiliation{
$^a$ Departamento de F\'{\i}sica, Funda\c{c}\~ao Universidade Federal do Rio Grande \\
Caixa Postal 474, CEP 96201-900, Rio Grande, RS, Brazil
\\
$^b$ Instituto de F\'{\i}sica e Matem\'atica, Universidade Federal de
Pelotas
\\
Caixa Postal 354, CEP 96010-090, Pelotas, RS, Brazil\\
}

\begin{abstract}
In this paper we estimate the influence of nonlinear gluon evolution
in the production of prompt photons at the LHC pp collider. We
assume  the validity of collinear factorization and consider the
EHKQS parton distributions, which are solutions of the GLR-MQ
evolution equations and describe quite well the DESY $ep$ HERA data,
as input in our calculations. We find that both single and double
photon production are enhanced for low-$p_T$ photons and central
rapidities, while this effect is absent for the high-$p_T$ photons.
The implications of this effect for the Quark-Gluon Plasma searches
and for the QCD background to Higgs are also discussed.

\end{abstract}

\maketitle

Prompt photons at high energies has provided a direct probe of the
dynamics of the strong interactions \cite{owens}. Usually photons
are called ¨prompt¨ if they are coupled to the interacting quarks.
In the framework of QCD the dominant production mechanism for the
prompt photons at Tevatron and LHC colliders is the Compton
scattering $q + g \rightarrow \gamma + q$ (See, e.g. Refs.
\cite{aurenche,kumar}). It is clear that the cross section of such
process is sensitive to the gluon distribution in the proton or
nuclei \cite{vogelsang}. In particular, LHC probes the gluon
distribution at $x_g \approx 10^{-5}$ \cite{kumar}. Usually the
parton distributions in a proton are described by the DGLAP
evolution equations \cite{dglap} and the cross section is calculated
using collinear factorization. More recently, the
$k_T$-factorization, which takes into account effects of finite
virtualities and transverse momenta of the incoming partons, has
been applied to calculate the prompt photon production cross section
\cite{lipatov,schurek}. In this case, the cross section is given in
terms of the unintegrated parton distributions, which are solution
of the BFKL or CCFM evolution equations, for instance. However, at
high energies (small-$x$), these linear evolution equations predict
a strong growth of the gluon distributions, which implies a large
density of gluons in this regime and the violation of the unitarity.
Consequently, new dynamical effects associated to the unitarity
corrections are expected to stop its further growth (For recent
reviews see Ref. \cite{hdqcd}).

About 24 years ago, Gribov, Levin and Ryskin (GLR) \cite{glr},
followed by Mueller and Qiu (MQ) \cite{qiu}, performed a detailed
study of the small-$x$ regime  and argued that at large densities
the physical processes of interaction and recombination of partons
become important and should be considered in the QCD evolution,
implying a nonlinear evolution equation. In the last two decades,
the solution and possible generalizations of the GLR-MQ equation
have been studied in great detail (See e.g. Ref. \cite{hdqcd}).
Currently, one believe that the small-$x$ gluons in a hadron
wavefunction should form a color glass condensate (CGC) which is
described by an infinite hierarchy of coupled evolution equations
for the correlators of Wilson lines \cite{cgc}. In the absence of
correlations, the first equation in the Balitsky-JIMWLK hierarchy
decouples and is then equivalent to the equation derived
independently by Kovchegov \cite{kov}. The resulting BK equation is
almost equivalent to the GLR-MQ equation for the unintegrated gluon
distribution, with the solutions presenting similar characteristics
\cite{kutak}. Experimentally, there are strong evidences of the
nonlinear (saturation) effects at DESY-HERA. In particular, the DESY
$ep$ HERA data in the small-$x$ and low-$Q^2$ region can be
successfully described in terms of saturation models
\cite{satmodels}, with the measured cross sections presenting the
geometric scaling property \cite{geosca}. Moreover, in Ref.
\cite{ehkqs} the authors have shown that adding nonlinear
corrections to the evolution, based on gluon recombination as
derived in \cite{glr,qiu}, can improve the overall leading order
(LO) fits to the HERA DIS data.  { Its main results are that when
nonlinear terms are included, the resulting gluon distribution is
reduced with respect to the solution of the LO BFKL equation and the
slowing of the $Q^2$ evolution leads to an enhancement of the
small-$x$ gluon distribution at $Q^2 \le 10$ GeV$^2$ relative to the
LO DGLAP gluon distributions. It implies similar enhancements in the
cross sections which are strongly dependent on the gluon
distribution with respect those calculated using the solution of the
LO DGLAP evolution equation}. In \cite{hq} the charm quark
production has been estimated at LHC energies and a substantial
enhancement was predicted. As the prompt photon production cross
section is directly dependent on the behavior of the gluon
distribution, one can expect a similar enhancement of the cross
section.

Our goal in this paper is estimate the magnitude of this enhancement
and determine the kinematic region where this effect should be more
important. We consider prompt photon production via the QCD
processes $q + \bar{q} \rightarrow \gamma + g$ and $q + g
\rightarrow \gamma + q$ in order to investigate by which amount
nonlinear effects contribute to the hard process. Our main focus is
the production of prompt photons in the kinematical region
accessible at CERN LHC energy. Moreover, as a by product, we also
estimate the contribution of these effects for the double photon
production, which constitutes an important QCD background to the
Higgs (h)  production chain $pp \rightarrow h X \rightarrow \gamma
\gamma X$.

\begin{figure}[t]
\includegraphics[scale=0.35]{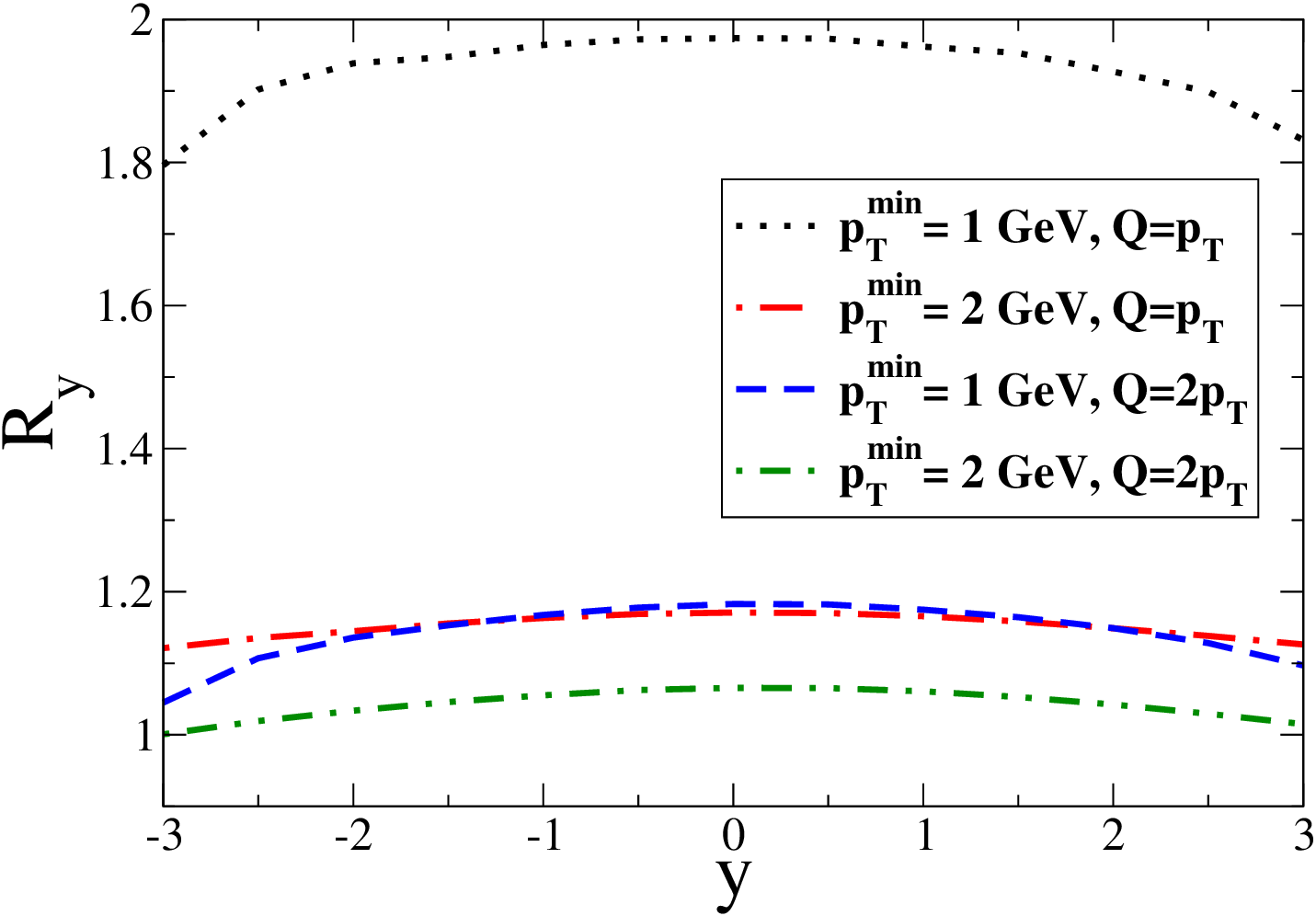}
\caption{(color online) Rapidity dependence of the ratio $R_{y}$ for
two different choices of the hard scale $Q$ and minimum transverse
momentum $p_T^{min}$. } \label{fig:1}
\end{figure}

Lets now present a brief review of the prompt photon production. Two
types of processes contribute  to the cross section: the so-called
direct piece, where the  photon is emitted via a point like coupling
to a quark, and the fragmentation piece, in which the photon
originates from the fragmentation of a final state parton. As the
second component  can be almost completely reduced by isolation
criterion used in the experimental data analysis, we focus our study
only in the direct component, which provide a clean probe of the
hard scattering dynamics. In this case, we have that the prompt
photon production cross section is given by \cite{owens}
\begin{eqnarray}
\frac{d\sigma_{pp\rightarrow \gamma X}}{dydp_T^2} = \sum_{i,j,k}
\int_{x_{1}^{min}}^1 dx_1 f_i(x_1,Q^2) f_j(x_2,Q^2) \nonumber \\
\frac{x_1x_2}{2x_1-x_Te^y}
 \frac{d \hat{\sigma}_{ij\rightarrow \gamma k}}{d\hat{t}} (Q^2,x_1,x_2) \,\,,
\label{cslo}
\end{eqnarray}
where $x_T=2p_T/\sqrt{s}$, $y$ and $p_T$ are the rapidity and
transverse momentum of the produced photon, $f_i (x,Q^2)$ are the
parton densities, $x_1$ and $x_2$ are the momentum fractions of the
partons involved in the hard process. In this case we have that
$x_2=\frac{x_1x_Te^{-y}}{2x_1-x_Te^y}$ and
$x_{1}^{min}=\frac{x_Te^{y}}{2-x_Te^{-y} }$. $Q^2$ is the hard scale
and $\frac{d\hat{\sigma}}{d\hat{t}}$ are the partonic cross
sections, which are perturbatively calculable \cite{field}. The
contributing LO subprocesses are $qg \rightarrow q \gamma$
(Compton), $q \bar q \rightarrow g \gamma$ (annihilation), followed
by the subdominant diagrams $q \bar q  \rightarrow \gamma \gamma$
(pure EM), $g g \rightarrow \gamma \gamma$ and $g g  \rightarrow g
\gamma$. The correspondent LO matrix elements and partonic cross
sections can be found in Refs. \cite{owens,field}. Using Eq.
(\ref{cslo}) to calculate the cross section, we implicity are
assuming the validity of the collinear factorization. It is
important to emphasize that factorization breaking is predicted by
CGC physics in several processes (See e.g. \cite{facbreaking}). In
what follows we calculate the prompt photon production cross section
considering the EHKQS \cite{ehkqs} and CTEQ6 \cite{cteq6}  sets for
the parton distributions and estimate the ratio among these
predictions in order to quantify the magnitude of the enhancement
expected at LHC. We analyze the rapidity and transverse momentum
dependence of this ratio. As the EHKQS sets are only evolved to
leading order (LO), we work the cross sections also at leading
order. It implies that our predictions should be considered an upper
bound, since the NLO small-$x$ gluon distributions are typically
reduced relative to LO, implying a smaller enhancement.

\begin{figure}[t]
\includegraphics[scale=0.35]{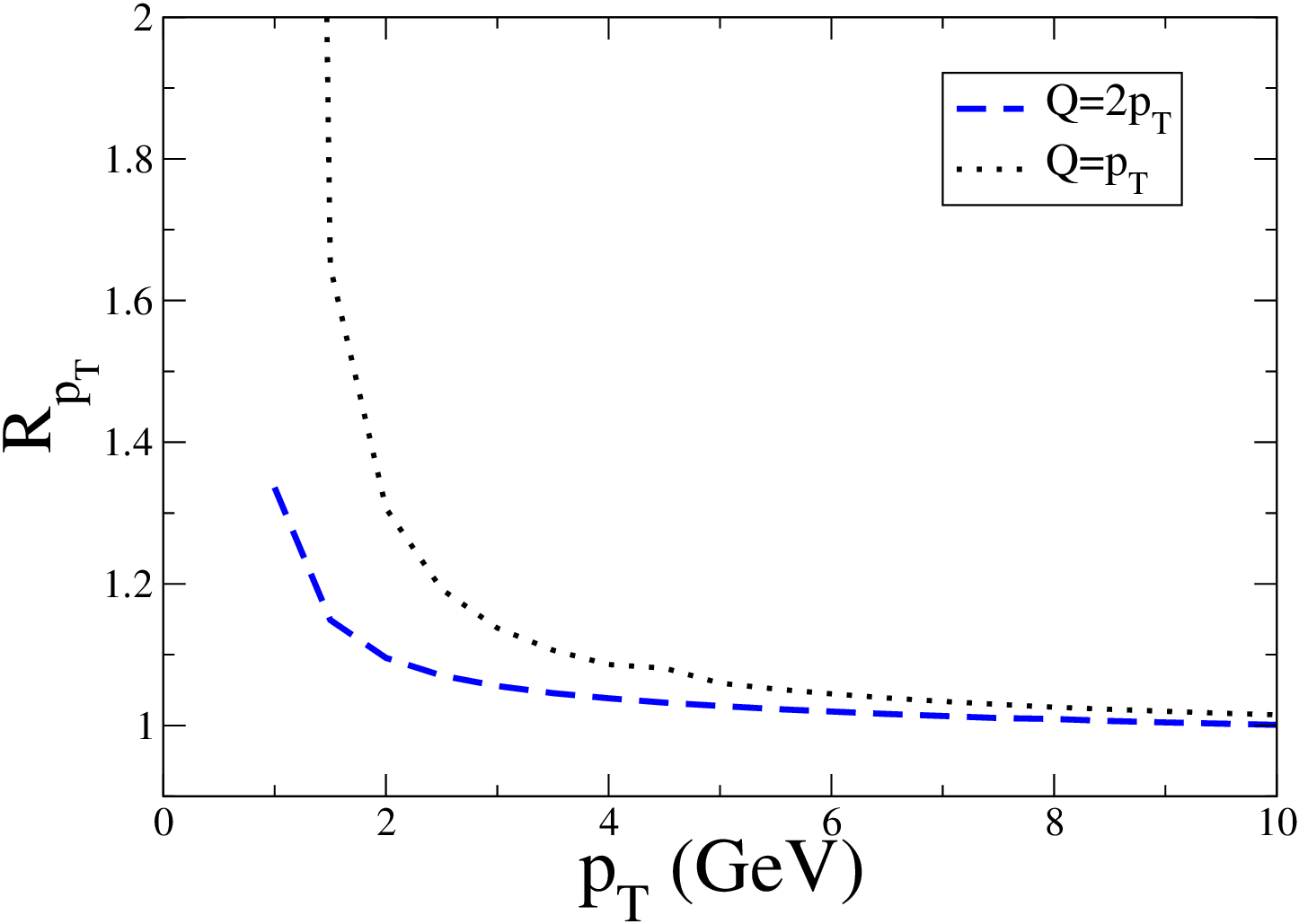}
\caption{(color online) Transverse momentum dependence of the ration
$R_{p_T}$ for two different choices of the hard scale $Q$. }
\label{fig:2}
\end{figure}

In Fig. \ref{fig:1} we present our estimates for the rapidity
dependence of the ratio $R_{y}$ defined by
\begin{eqnarray}
R_{y}  \equiv { \frac{d\sigma (EHKQS)}{dy} } / {\frac{d\sigma
(CTEQ6)}{dy}} \,\,.
\end{eqnarray}
We consider two different values for the factorization and
renormalization scales: $Q = p_T$ and $Q = 2 p_T$, where $p_T$ is
the photon transverse momentum. Moreover, in order to calculate the
differential cross section $d\sigma / dy$ we have assumed two
distinct values for the minimum photon transverse momentum:
$p_T^{min} = 1$ and 2 GeV. The results show an enhancement of low-$p_T$ photons
-
the lower $p_T^{min}$ the more significant enhancement of prompt
photon production. We also notice a larger enhancement for central
rapidities. The choice of the hard scale also affects the magnitude
of the nonlinear effects, by choosing $Q = p_T$ one has a bigger
enhancement. In particular,  for $p_T^{min} = 1$ GeV and $Q = p_T$,
we predict a factor of two for the enhancement of the rapidity
distribution associated to the nonlinear effects in the central
rapidity region.

\begin{figure}[t]
%\vspace{.5cm}
\includegraphics[scale=0.35]{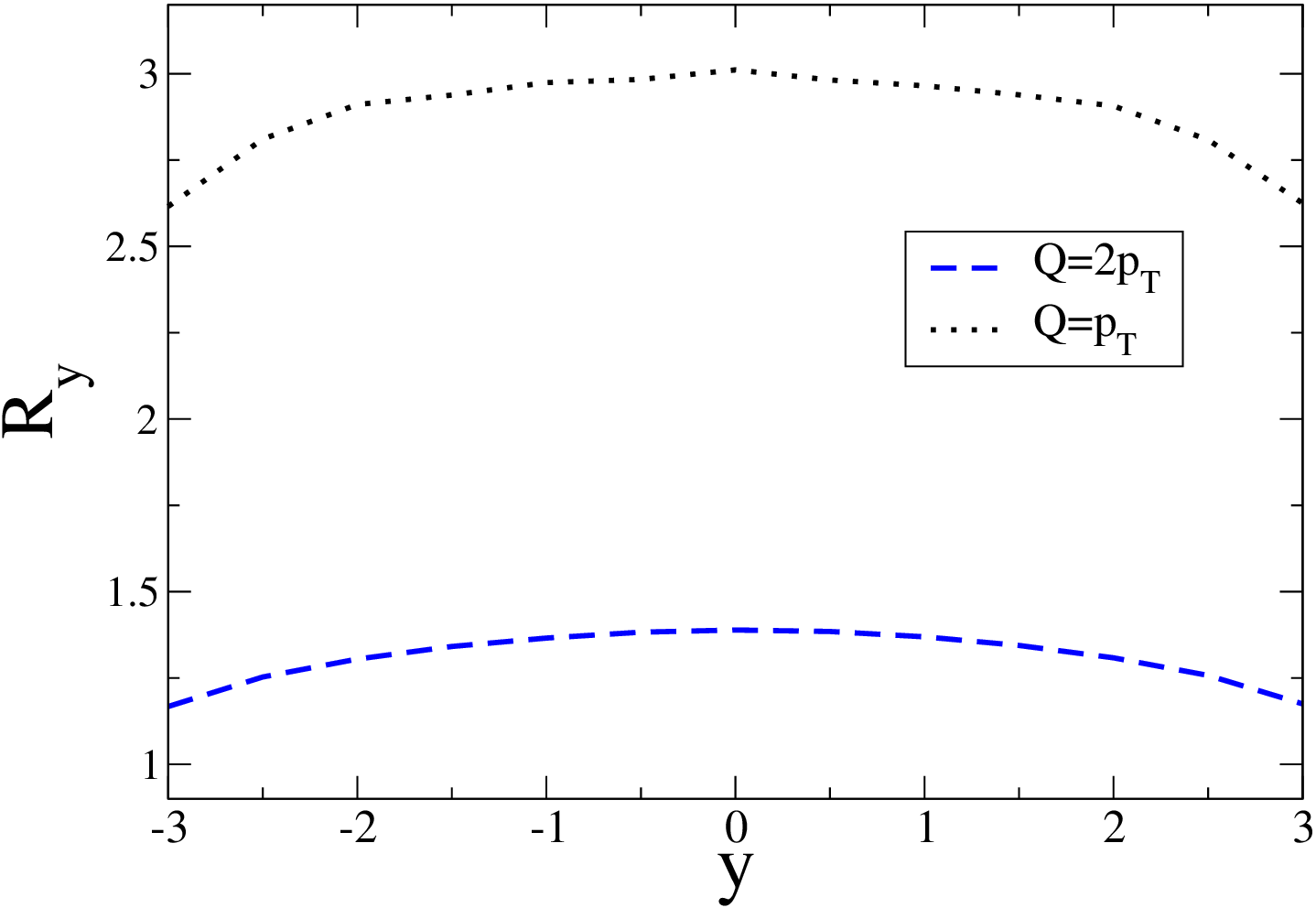}
\caption{(color online) Rapidity dependence of the ratio $R_{y}$
calculated considering only the $q \bar{q} \rightarrow \gamma
\gamma$ and $gg \rightarrow \gamma \gamma$ subprocesses.}
\label{fig:3}
\end{figure}

In Fig. \ref{fig:2} we present our estimates for the transverse
momentum dependence of the ratio $R_{p_T}$ defined by
\begin{eqnarray}
R_{p_T}  \equiv { \frac{d\sigma (EHKQS)}{d^2 p_T} } / {\frac{d\sigma
(CTEQ6)}{d^2 p_T}} \,\,.
\end{eqnarray}
Similarly to our previous analysis we also consider two different
choices for the scale $Q$. Moreover, we have integrated the rapidity
distribution in the range $|y| < 3$, which will be studied in LHC.
We have that an enhancement is predicted in the region of small
transverse momentum $p_T < 10$ GeV. At larger scales  the
nonlinearities die out since these terms are proportional to
$1/Q^2$. Consequently, the EHKQS gluons become similar to the CTEQ6
one, so that the enhancement disappears at large $p_T$. This
behavior is also expected in charm production \cite{hq}. From the
Fig. \ref{fig:2} we can see that the nonlinear effects imply an
enhancement of 20 $\%$ in the small-$p_T$ region if we assume $Q = 2
p_T$. On the other hand, for $Q =  p_T$ this enhancement can be
larger than 100 $\%$. The enhancement in the low range of transverse
momenta has important implications, mainly if we  remember that the
magnitude of prompt photon production in proton-proton collisions is
used as baseline to estimate nuclear medium effects in
nucleus-nucleus collisions \cite{hardprobes}. In particular, a
signature of quark-gluon plasma formation is the thermal photon
production, which manifests by an enhancement in the inclusive
photon spectrum at $p_T$ values below $\approx 15$ GeV at LHC (For
recent reviews see Refs. \cite{thoma1,Stankus}). The pre-requisite
for the extraction of the thermal signal is a precise control of the
photon production rate in proton-proton collisions, which can be
strongly modified by the nonlinear effects, as demonstrated above.
Consequently, our results indicate that a detailed study of the
prompt photon production in $pp$ collisions at LHC is necessary
before to consider the thermal photon production as a precise signal
of QGP formation.

One can analyze the contribution of the nonlinear effects for the
diphoton production, characterized by the subprocesses $q \bar{q}
\rightarrow \gamma \gamma$ and $gg \rightarrow \gamma \gamma$, which
have been included in our calculations. Although the gluon-gluon
initiated process is not the dominant one, it contributes
significantly for the $\gamma \gamma$ cross section. This may imply
in bigger nonlinear effects, since both initial state small-$x$
gluons should be amplified in the region $Q^2 \le 10$ GeV$^2$. In
Fig. \ref{fig:3} we present our results for the rapidity
dependence of the ratio $R_{y}$, calculated including only the $q
\bar{q}$ and $gg$ contributions. We consider the same two choices of
factorization and renormalization scales as before: $Q = p_T$ and $Q
= 2 p_T$, and we have assumed the value of $p_T^{min} = 1$ GeV for
the minimum photon transverse momentum. We can see that the
enhancement due to nonlinear effects is more significant here than
for single photon production - probably due the the relative
importance of the two gluon initiated contribution. In fact, if we
had considered only the $gg \rightarrow \gamma \gamma$
 and neglected the $q \bar{q} \rightarrow \gamma \gamma$ contribution (just for theoretical
 study), the $R_{y}$ ratio would increase to almost 5 (1.5) for the $Q = p_T$ ($Q = 2
 p_T$) choice. Anyway, in the full LO contribution some of this increased enhancement
 remains.

\begin{figure}[t]
\includegraphics[scale=0.35]{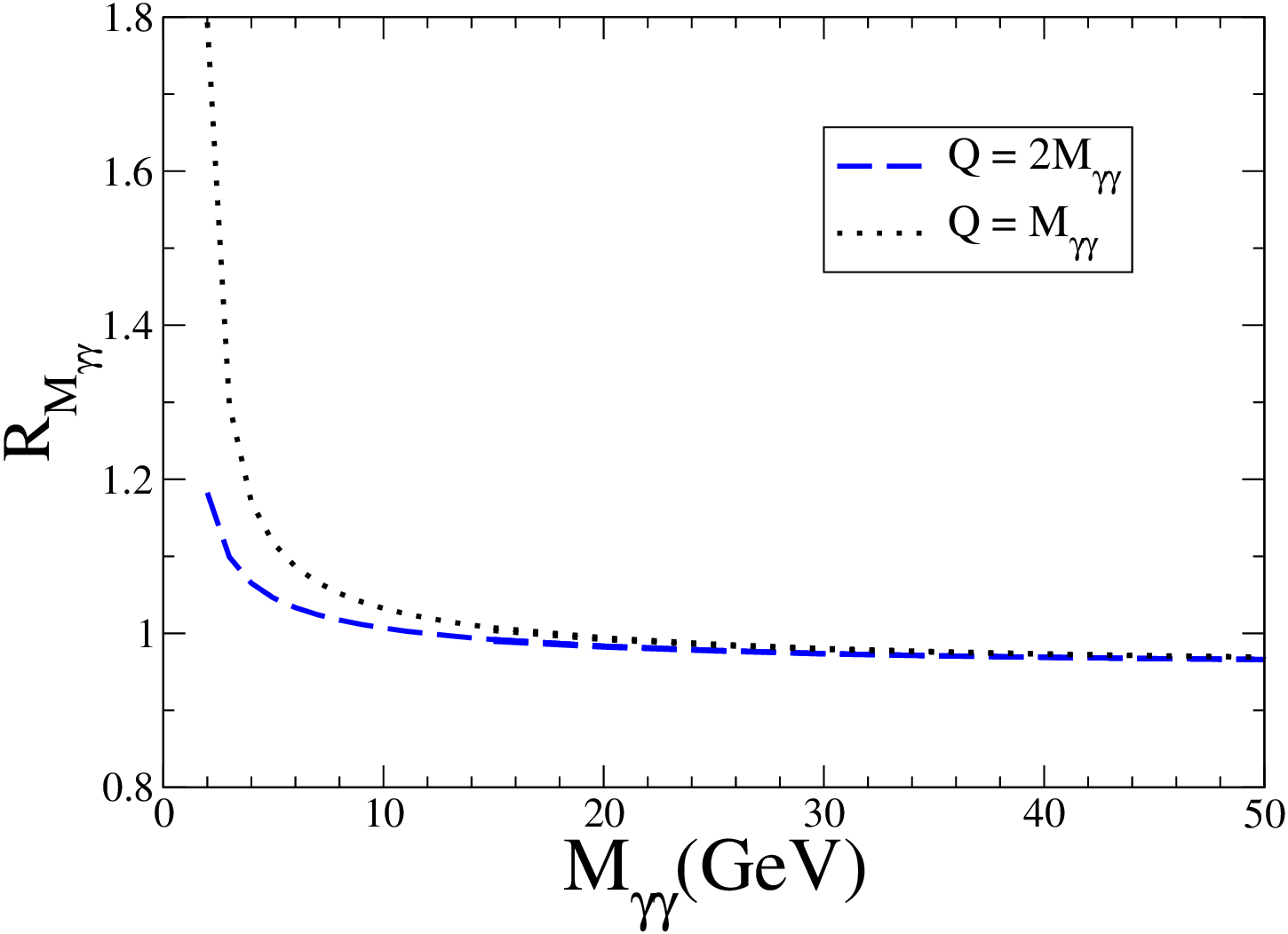}
\caption{(color online) Diphoton invariant mass dependence of the
ratio $R_{M_{\gamma \gamma}}$.} \label{fig:4}
\end{figure}

This result motivates a more detailed analysis of diphoton
production, which is considered the main background for Higgs
production via gluon fusion (See e.g. Ref. \cite{ingelman}). In
terms of the diphoton invariant mass $M_{\gamma \gamma}$ and photon
rapidities $y_1$ and $y_2$, the cross section can be written as
\cite{owens}
\begin{eqnarray}
\frac{d\sigma_{pp\rightarrow \gamma \gamma} }{dy_1dy_2dM_{\gamma \gamma}}=
\sum_{ij}x_1f_i(x_1,Q^2)x_2f_j(x_2,Q^2)\nonumber\\
\frac{M_{\gamma \gamma}}{1+\cosh (y_1-y_2)}
\frac{d \hat{\sigma}_{ij\rightarrow \gamma \gamma}}{d\hat{t}} (Q^2,x_1,x_2)\,\,,
\end{eqnarray}
where $x_1=\frac{p_T}{\sqrt s}[e^{y_1}+e^{y_2}]$ and $x_2=\frac{p_T}{\sqrt
s}[e^{-y_1}+e^{-y_2}]$.
In particular, we have calculated the
diphoton invariant mass distribution, $d \sigma/dM_{\gamma \gamma}$,
and estimated the ratio $R_{M_{\gamma \gamma}}$ defined by
\begin{eqnarray}
R_{M_{\gamma \gamma}}  \equiv { \frac{d\sigma (EHKQS)}{dM_{\gamma
\gamma}} } / {\frac{d\sigma (CTEQ6)}{dM_{\gamma \gamma}}} \,\,.
\end{eqnarray}
In Fig. \ref{fig:4} we present our predictions for the behavior of
this ratio in the kinematical range of $M_{\gamma \gamma} < 50$ GeV
considering two distinct values of the hard scale $Q$. We have that
in this range an enhancement is predicted, mainly at small values of
 the factorization scale $Q$ and diphoton invariant mass. On the other hand, for larger values
of $M_{\gamma \gamma}$ the nonlinear effects can be disregarded. Since the typical $p_T$ of
photons needed to produce a Higgs would be around 50 GeV each, the
nonlinear effects here considered will not be important for the QCD
background to Higgs (For a discussion of nonlinear effects in Higgs
production see e.g. Ref. \cite{motyka}). On the other hand, the
enhancement in the low invariant mass region is predicted so the
nonlinear contributions could be tested in this range.

In summary, in this paper we have investigated the prompt photon
production considering the collinear factorization and the EHKQS
parton distributions, which are solutions of the GLR-MQ evolution
equations and describe quite well the $ep$ HERA data. Our results
demonstrate that the nonlinear effects implies a large enhancement
of the cross section for single and double photon production, which
could be tested at LHC. In particular, the magnitude of the single
photon production at small-$p_T$ should be constrained before using
thermal photon production as a signature of the QGP formation. On
the other hand, the nonlinear effects can be disregarded in the
kinematical range where double photon production is an important
background for Higgs production.

\begin{acknowledgments}
This work was  partially financed by the Brazilian funding agencies
FAPERGS and CNPq.
\end{acknowledgments}

\end{document}